\documentclass[letterpaper,twocolumn,10pt]{article}
\usepackage{usenix,epsfig,endnotes}

\usepackage[normalem]{ulem}
\usepackage{xspace}
\usepackage{latexsym}
\usepackage{graphicx}
\usepackage[hyphens]{url} 
\usepackage{verbatim} 

\usepackage[nameinlink]{cleveref}
\crefname{section}{\S}{\S\S}

\usepackage{cleveref}
\crefname{section}{�}{��}
\Crefname{section}{�}{��}

\usepackage{fancyhdr}
\newcommand{\etal}{\emph{et\,al.}\xspace}

\newcommand{\da}{DevAgent}

\begin{document}

\date{}


\title{
  Position: On Failure Diagnosis of the Storage Stack
}

%
\author{
{\rm Duo Zhang,  Om Rameshwar Gatla, Runzhou Han,   Mai Zheng} \\ 
 \textit{Iowa State University}\\
}
 

\maketitle



\subsection*{Abstract}
\vspace{-0.1in}
Diagnosing storage system failures is challenging even for 
professionals. 
One example is the ``When Solid State Drives Are Not That Solid''
incident occurred at Algolia data center, where Samsung SSDs were
mistakenly blamed for failures caused by a Linux kernel bug.
With the system complexity keeps increasing, 
such obscure failures will likely occur more often.

As one step to address the challenge,
we present our on-going efforts called X-Ray.
Different from traditional methods that 
focus on either the software or the hardware, 
X-Ray leverages virtualization to collects events across layers,
and correlates them to generate a correlation tree.
Moreover, by applying simple rules, 
X-Ray can highlight critical nodes  automatically. 
Preliminary results based on 5 failure cases 
shows that X-Ray can 
effectively narrow down the search space for failures.

\vspace{-0.1in}
\section{Motivation}
\label{sec:intro}
\vspace{-0.1in}

The storage stack is
witnessing a sea-change 
driven by the 
advances in non-volatile memory (NVM)
 technologies
~\cite{Grupp12, cai_date12, 
Grupp09, kurata_vlsi06, belgal_irps02, suh_jssc95, ong_vlsi93, 
brand_irps93,paging-pm}.
For example, 
flash-based solid state drives (SSDs) 
and persistent memories (PMs)
are replacing hard disk drives (HDDs) as the durable device 
~\cite{hdd_ssd_pcshare,enterprise_ssd, meza_sigmetrics15, algolia_blog, flash_array,Intel-Dell-PM};
NVMe~\cite{nvme} and CXL~\cite{cxl} are redefining the host-device interface; 
\textit{blk-mq}~\cite{Bjorling_systor13} alleviates the single queue 
and lock contention bottleneck 
at the block I/O layer;
the SCSI subsystem and the Ext4 file system, 
which have been tuned for HDDs for decades,
are also being adapted for NVM
(e.g., scsi-mq~\cite{lwn-scsi-mq, ornl-scsi-mq, van2015-scsi-mq} 
and DAX~\cite{Ext4_DAX});
in addition, various NVM-oriented new designs/optimizations
 have been proposed  (e.g., F2FS~\cite{f2fs}, NOVA~\cite{nova}, 
 Kevlar~\cite{software-wear-management-fast19}), some of which require cohesive
modifications throughout the storage stack (e.g., the TRIM support~\cite{trim}).


The new systems generally offer higher performance.
However, 
as a disruptive technology, the NVM-based
components
have to co-exist with the traditional storage ecosystem, 
which is notoriously complex and difficult to 
get right despite decades of efforts~\cite{Yang-OSDI06-EXPLODE, iron05, Lu2014, stallman2002debugging}.
Compared with the performance gain,
 the implication on system reliability is much less 
studied or understood.

One real example is the ``When Solid-State Drives Are Not That Solid'' incident 
occurred in Algolia data center~\cite{algolia_blog}, where
a random subset of SSD-based servers crashed and corrupted files for unknown reasons.
The developers ``spent a big portion of two weeks just isolating
machines and restoring data 
as quickly as possible''.  
After trying to diagnose almost all  software  in the  stack 
(e.g., Ext4, \texttt{mdadm}~\cite{mdadm}), 
and switching SSDs from different vendors,
they
finally (mistakenly) concluded that it was Samsung's SSDs to blame.
Samsung's SSDs were criticized and blacklisted,
until one month later Samsung engineers found that
it was a TRIM-related Linux kernel bug that caused the failure~\cite{algolia-samsung-patch}.


As another example, 
Zheng \etal{}  studied the behavior of SSDs 
under power fault~\cite{Zheng-FAST13-SSD}. 
The testing framework bypassed the file system,
but relied on the block I/O layer to apply workloads
and check the behavior of devices.
In other words,
the SSDs were essentially evaluated together with the block I/O layer.
Their initial experiments were performed on Linux kernel v2.6.32,
and eight out of fifteen SSDs exhibited 
a symptom called ``serialization errors''~\cite{Zheng-FAST13-SSD}. 
However, 
in their follow-up work where 
similar experiments were conducted on a newer kernel (v3.16.0)~\cite{Zheng-TOCS16-SSD}, 
the authors observed that the failure symptoms on some SSDs 
changed significantly (see Table \ref{tab:two_kernels_avg_err}, 
adapted from ~\cite{Zheng-TOCS16-SSD}).
It was eventually confirmed that the different symptoms was caused 
by a sync-related Linux kernel bug~\cite{Zheng-TOCS16-SSD}.

\begin{table}[tb]
  \begin{center}
  \begin{tabular}{c|c|c|c}
  \hline
  OS (Kernel) & SSD-1   &SSD-2     &SSD-3 \\
  \hline
  Debian 6.0 (2.6.32)       & 317    & 27      & 0 \\
  Ubuntu 14.04 (3.16)   &  88    & 1        &  0\\
  \hline
  \end{tabular}
  \end{center}
  \vspace{-3ex} 
  \caption{{\bf 
  SSDs exhibit different symptoms when tested 
  on different OSes.  
  }
  {\it 
  Each cell shows the average number of errors. 
  Reported by~\cite{Zheng-TOCS16-SSD}.
  }
  }
  \label{tab:two_kernels_avg_err}
  \vspace{-0.2in}
  \end{table}

One commonality of the  two cases above is that people
try to infer the behavior of storage 
devices \textit{indirectly} through the operating system (OS) kernel, 
and they tend to believe that the kernel is correct. 
This is natural in practice because  users typically  
have to access 
storage devices 
with the help of the kernel,
and they usually do not have the luxury of inspecting the 
 device behavior
 {directly}.
Also, NVM devices are relatively young compared with
 the long history of 
the OS kernel, 
so they might seem less trustable. 
We call such common practice as a \textit{top-down} approach. 


Nevertheless, both cases show that the OS kernel may play a role 
in causing system failures, while the device may be innocent. 
More strangely, in both cases,
different devices seem to have different sensitivity to the kernel bug, 
and some devices may even ``tolerate'' the kernel bug. 
For example, no failure was observed on Intel SSDs in the Algolia case~\cite{algolia_blog},
and the SSD-3 in Table~\ref{tab:two_kernels_avg_err} 
never exhibited any serialization errors
 in Zheng \etal{}'s experiments~\cite{Zheng-FAST13-SSD}.
Since switching devices is one simple and common strategy
to identify device issues when diagnosing system failures, 
the different sensitivity of devices
to the software bugs can easily drive the investigation to the wrong 
direction, wasting human efforts and resulting in wrong conclusions,
as manifested in the two cases above.

In fact, similar confusing and debatable failures 
are not uncommon today~\cite{trim-zfs,trim-libata,trim-mSATA,HP_warn}.
With the trend of storage devices becoming more capable 
and more special features are being exposed to  the host-side software~\cite{BeyondBlockIO-HPCA,2BSSD,FlatFlash},
the interaction between hardware and  software is expected to 
be more complex.
Consequently, 
analyzing storage system failures 
solely based on the existing 
{top-down} approach will likely become more problematic. 
In other words,  new methodologies for 
diagnosing failures of the storage stack are much needed.

The rest of the  paper is organized as follows: 
First, we discuss the limitations of existing efforts (\S\ref{sec:why});
Next, we introduce our idea (\S\ref{sec:design})
and the  preliminary results (\S\ref{sec:experiments}); 
Finally, we describe other related work (\S\ref{sec:related})
and conclude with the discussion topics section (\S\ref{sec:discussion}).


\vspace{-0.1in}
\section{Why Existing Efforts Are Not Enough}
\label{sec:why}
\vspace{-0.1in}
In this section, we discuss two  
groups of existing efforts 
that may 
alleviate the challenge of diagnosing storage stack failures to some extent.
We defer the discussion of other related work
(e.g., diagnosing performance issues and distributed systems)
to \S\ref{sec:related}. 

\vspace{-0.1in}
\subsection{Testing the Storage Software  Stack}
\label{sec:whynot_test}
\vspace{-0.05in}
Great  efforts have been made to {test} the storage software  in the stack~\cite{Yang-OSDI06-EXPLODE,Changwoo-SOSP15-CrosscheckingFS,vjay-osdi18-blackboxtesting,Zheng-OSDI14-DB},
with the goal of exposing bugs that could lead to failures.
For example, EXPLODE~\cite{Yang-OSDI06-EXPLODE} and B$^3$~\cite{vjay-osdi18-blackboxtesting}  
apply fault injections 
to detect crash-consistency bugs in file systems.

However, testing tools are generally not suitable for diagnosing  system failures 
because they typically require a well-controlled  environment (e.g., a highly customized kernel~\cite{Yang-OSDI06-EXPLODE,vjay-osdi18-blackboxtesting}), 
which may be substantially different from 
the storage stack that need to be diagnosed.
\vspace{-0.1in}
\subsection{Practical Diagnosing Tools}
\label{sec:whynot_existingtools}
\vspace{-0.05in}
To some extent, failure diagnosis 
is the reverse process of fault injection testing. 
Due to the importance, 
many practical tools have been built, including the following:

\smallskip
\noindent
{\bf Debuggers} ~\cite{gdb,kdb,kgdb}
  are the \textit{de facto} way to diagnose system failures.
They  usually support fine-grained manual inspection  
 (e.g., set breakpoints, check memory bytes).
 However, significant human efforts are needed to harness the power and diagnose the storage stack. The  manual effort required will keep increasing as the software becomes more complex. 
 Also, 
 these tools typically cannot collect the device information directly.

\smallskip
\noindent
{\bf Software Tracers} 
~\cite{GoogelXRay,kprobes,Cantrill2004,SOSP11-WindowsFay}
  can collect various events from a target system
to help understand the behavior.  
However, similar to debuggers, they focus on host-side  events only, 
and  usually do not have automation support for 
 failure inspection. 

\smallskip
\noindent
{\bf Bus Analyzers} ~\cite{busanalyzers,ananlyer2}
are hardware equipments that 
can capture  
 the communication data between a host system and a device, 
which are particularly useful for  analyzing the device behavior.
However, since they only report bus-level information,
they cannot help much on understanding system-level behaviors.

Note that both debuggers and software tracers represent the 
traditional top-down diagnosis approach. On the other hand,
bus analyzers have been used to diagnose 
some of the most obscure failures that involved
 host-device interactions~\cite{host1,host2},
but they are not as convenient as the software  tools. 
\vspace{-0.1in}
\section{X-Ray: A Cross-Layer Approach}
\vspace{-0.1in}
\label{sec:design}

Our goal is to help practitioners to narrow down the 
the root causes of storage system failures quickly. 
To this end, we are exploring a framework called \texttt{X-Ray},
which is expected to have the following key features: 

\begin{figure*}[tb]
    \begin{center}
           \includegraphics[width=5.8in]{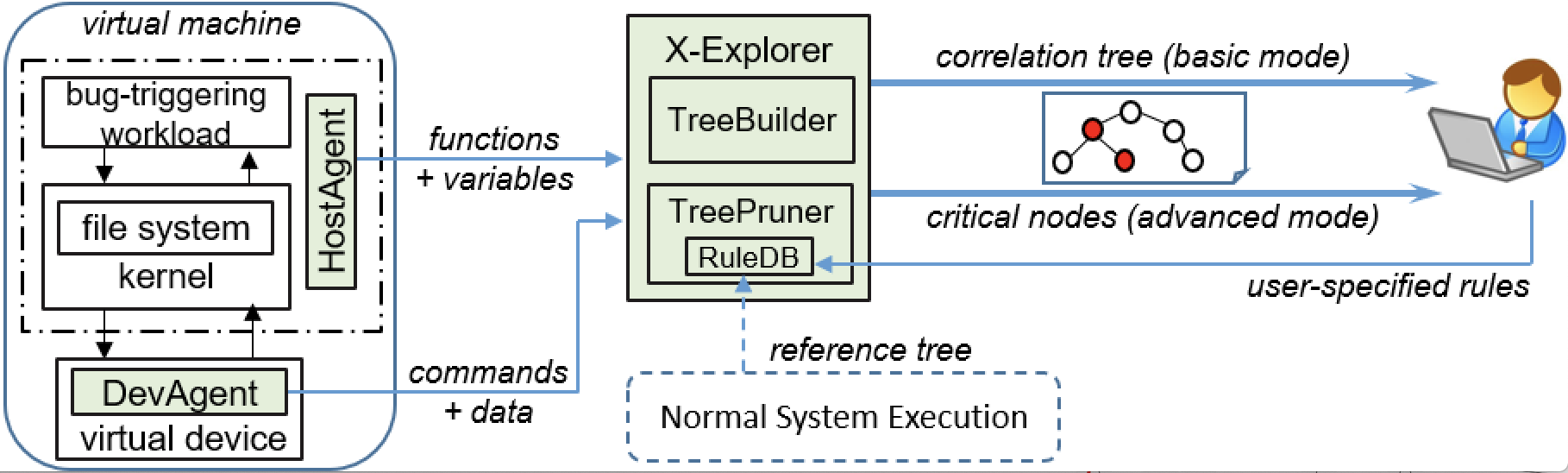}
    \end{center}
    \vspace{-0.3in}
    \caption{
        {\bf The X-Ray Approach}. 
       {\it The target  software stack is hosted in a virtual machine;
   {DevAgent},  {HostAgent}, and {X-Explorer}
   are the three main components; the basic mode visualizes a correlation tree 
   for inspection; the advanced mode highlights
   critical nodes based on rules.
    }
    }
    \vspace{-0.2in}
    \label{fig:overview}
\end{figure*}

\begin{itemize}
    \vspace{-0.05in}
    \item  {\bf Full stack:} many critical operations (e.g., sync, TRIM) 
    require cohesive support at both device and host sides; 
    inspecting only one side (and assuming the other side is correct) is fundamentally limited;
    
    \vspace{-0.05in} 
    \item {\bf Isolation:} the device information should be collected 
    without relying on the host-side software (which may be problematic itself); 


    \vspace{-0.05in} 
    \item {\bf Usability:} 
    no special hardware or software modification is needed; 
    manual inspection should be reduced as much as possible.

  
    

\end{itemize}

\vspace{-0.2in}
\subsection{Overview}
\label{sec:overview}
\vspace{-0.05in}


Figure~\ref{fig:overview} shows an overview of the \texttt{X-Ray} framework,
which includes three major components: \textit{DevAgent},  \textit{HostAgent}, and \textit{X-Explorer}.

First, we notice that the virtualization technology is mature enough to
support unmodified OS today~\cite{VMware,VirtualBox,PWorkstation}.
Moreover, 
recent research efforts  have enabled emulating sophisticated storage devices in a virtual machine (VM), 
including SATA SSDs (e.g., VSSIM~\cite{vissim}) and NVMe SSDs (e.g., FEMU~\cite{femu}).
Therefore, we can leverage virtualization to support cross-layer analysis
with high fidelity and no hardware dependence.  

Specifically, we host the target storage software stack in a 
QEMU VM~\cite{qemu}. At the virtual device layer, 
the DevAgent (\S\ref{sec:devagent})  monitors  the commands (e.g.,
SCSI, NVMe) transferred from the kernel under the bug-triggering workload.
Optionally, 
the associated data (i.e., bits transferred by  commands) can be recorded too.


Meanwhile, to understand the high-level semantics of the system activities,
the HostAgent (\S\ref{sec:hostagent}) monitors the function invocations 
throughout the software stack (e.g., system calls, kernel internal functions),
and records them with timestamps at the host side. 
Optionally, key variables
 may be recorded too with additional overhead.

 The X-Explorer (\S\ref{sec:xexplorer}) helps to 
 diagnose  the  system behavior 
in two modes:
(1) the basic mode visualizes a correlation tree of cross-layer events 
 for inspection; 
 (2) the advanced mode highlights critical events 
 based on rules, which 
 can be either
 specified by the user or derived from a normal system execution.

\vspace{-0.1in}
\subsection{DevAgent}
\label{sec:devagent}
\vspace{-0.05in}

The device-level information is helpful
because storage  failures are often 
related to the persistent states, 
and changing persistent states (in)correctly requires  (in)correct device command sequences.
The DevAgent records the information 
in a command log directly 
 without any dependency on the host-side kernel 
 (which might be  buggy itself),
similar to the  bus analyzer~\cite{busanalyzers}.

\smallskip
\noindent
\textbf{SCSI Device.} 
The Linux kernel communicates with a SCSI device by sending 
Command Descriptor Blocks (CDBs) over the bus.
 QEMU maintains a \texttt{struct SCSICommand} for
 each SCSI command,
 which contains a 16-byte buffer (\texttt{SCSICommand->buf})  holding the CDB. 
 Every SCSI command type is identified by the opcode at the beginning of the CDB, 
 and the size of CDB  is determined by the opcode. 
 For example, the CDB for
 the WRITE\_10
 command is represented by the first 10 bytes of the buffer. 
For simplicity, we always  transfer 16 bytes from the buffer to the command log and use the opcode 
 to identify  valid bytes.
QEMU classifies  SCSI commands into either 
Direct Memory Access (DMA) commands (e.g., READ\_10)
or Admin commands (e.g., VERIFY\_10), and both are handled in the same 
way in DevAgent since they share the same structure.



\smallskip
\noindent
\textbf{NVMe Device.} 
QEMU maintains a \texttt{struct NvmeCmd} for each NVMe command,
and emulates the \texttt{io\_uring}\cite{lwn-net-ringing-asynchronous, iouring-interface-patch} interface
 to transfer NVMe commands 
 to a NVMe device, 
The interface defines  two types of command queues:
 submission and completion. 
The submission queues are further classified into 
either I/O submission queue or Admin submission queue,
which are processed via 
\texttt{nvme\_process\_sq\_io} and \texttt{nvme\_process\_sq\_admin} 
in QEMU respectively.
The DevAgent intercepts both queues and 
records both I/O commands and Admin commands, similar to SCSI.

\begin{figure*}[tb]
    \begin{center}
           \includegraphics[width=6.4in]{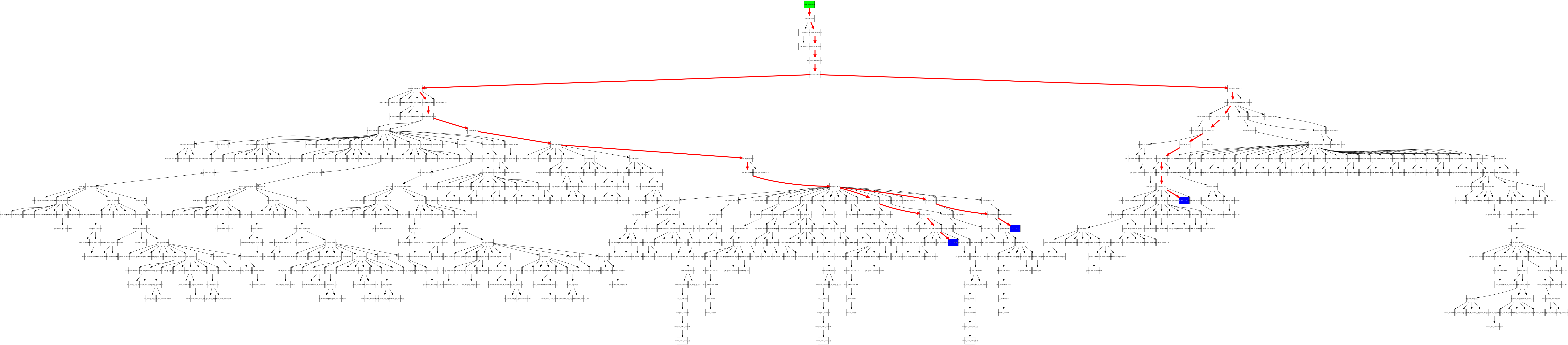}
    \end{center}
    \vspace{-0.3in}
    \caption{
        {\bf  A partial correlation tree.} 
       {\it 
       The tree includes one syscall (green), 704 kernel functions (white nodes), 
       and  3  device commands  (blue); 
       the critical path (red) is selected by a simple rule: all ancestors of the command nodes. 
    }
    }
    \vspace{-0.2in}
    \label{fig:crosslayergraph}
\end{figure*}

\vspace{-0.1in}
\subsection{HostAgent}
\label{sec:hostagent}
\vspace{-0.05in}

The HostAgent aims to track host-side events to
 help understand the high level semantics of  system activities.
     As mentioned in \S\ref{sec:why}, 
     many   tracers have been developed  with different tradeoffs~\cite{SoftwareTracers}.
   The current prototype of HostAgent is based on \texttt{ftrace}~\cite{ftrace},
   which has native support on Linux based on \texttt{kprobes}~\cite{kprobes}. 
     We select \texttt{ftrace}~\cite{ftrace} because  its convenient support on 
    tracing  caller-callee relationship. 
     When \texttt{CONFIG\_FUNCTION\_GRAPH\_TRACER} is defined, 
    the \texttt{ftrace\_graph\_call} routine will store the pointer to the parent 
     to a ring buffer at function return via the link register, 
     which is ideal for X-Ray.
     On the other hand, \texttt{ftrace} only records function execution time instead of
     the epoch time needed for synchronization with DevAgent events.
     To workaround the limitation, we modify the front end of \texttt{ftrace} 
     to record the epoch time at system calls, and calculates the epoch time 
     of kernel functions based on their execution time since the corresponding 
     system calls.
     Another issue we observe is that \texttt{ftrace} may miss executed 
     kernel functions.
     We are working on improving the completeness.

\vspace{-0.1in}
\subsection{X-Explorer}
\label{sec:xexplorer}
\vspace{-0.05in}
The events collected by DevAgent and HostAgent are valuable for diagnosis.
However, the quantity is usually too large for manual inspection.
Inspired by the visualization layer of other diagnosis tools~\cite{Bhatia-OSDI08,HotCloud09Mochi,XXXSOSP03-PerformanceDebugging,NSDI11-DiagnosingPerformance},
the X-Explorer visualizes the relationships among the events 
and highlights the critical ones. 


\vspace{-0.1in}
\subsubsection{TreeBuilder}
\label{sec:treebuilder}
\vspace{-0.05in}

The {TreeBuilder}
generates a {\it correlation tree} 
to represent the relationships
 among events in the storage stack.
The tree contains three types of nodes
based on the events from HostAgent and DevAgent:
(1) SYSCALL nodes represent the system calls invoked in the bug-triggering workload; 
 (2) KERNEL nodes represent the internal kernel functions involved;
(3) CMD nodes represent the commands observed at the device.

There are two types of edges in the tree: (1) the edges among SYSCALL 
 and KERNEL nodes represent function invocation relations 
 (i.e., parent and child);
 (2) the edges between CMD nodes and  other nodes represent  
 {\it close relations in terms of time}. 
In other words, the device-level events are 
correlated to the host-side events 
based on timestamps. 
While the idea is straightforward, 
we observe an out-of-order issue caused by  virtualization:
the HostAgent timestamp is collected within the VM,
 while the DevAgent timestamp is collected outside the VM;
the device commands may appear to occur before the corresponding system calls 
based on the raw timestamps. 
To workaround the issue, we set up an NTP server~\cite{NTP} at the DevAgent side
and perform NTP synchronization at the HostAgent Side.
We find that such NTP based synchronization may mitigate the timestamp gap 
to a great extent, as will be shown in \S\ref{sec:experiments}.
Another potential solution is to modify the dynamic binary translation
(DBT) layer of QEMU to minimize the latency.

\vspace{-0.1in}
\subsubsection{TreePruner}
\label{sec:treepruner}
\vspace{-0.05in}
The correlation tree  
is typically large due to the complexity of the storage stack.
Inspired by the rule-based diagnosis tools~\cite{Bhatia-OSDI08},
the {TreePruner} traverses the tree and highlights the {\it critical paths and nodes} 
(i.e., the paths and nodes of interest)
  automatically 
  based on a set of rules stored in the RuleDB,
  which can be either specified 
  by the user or  derived from a normal system execution.


\smallskip
\noindent
{\bf User-specified rules.} 
Users may specify expected relations among system events as rules.
For example,  the sync-family system calls 
(e.g., \texttt{sync}, \texttt{fsync}) should generate SYNC\_CACHE (SCSI) or FLUSH (NVMe) commands to the device, which is crucial for crash consistency; similarly, \texttt{blkdev\_fsync} should be triggered when calling \texttt{fsync} on a raw block device.
In addition, users may also specify simple rules to reduce the tree (e.g., all ancestor nodes of WRITE commands).  

Our current prototype hard-coded a few rules 
as tree traversal operations 
based on the failure cases we studied (\S\ref{sec:experiments}). 
We are exploring  more flexible 
interfaces 
(e.g., SQL-like~\cite{OSDI08-SQCK} or formula-based~\cite{ASPLOS16-Specifying}) to enable expressing more sophisticated rules.

\smallskip
\noindent
{\bf Normal system execution.} 
Failures are often tricky to reproduce 
 due to  different  environments
  (e.g., different kernel versions)
 ~\cite{diff_k}.
 In other words, 
failures may not always manifest even under the same bug-triggering workloads.
Based on this observation,  and inspired by delta debugging~\cite{delta_debug,hierarchicaldeltadebugging},
we may leverage   a normal system execution 
 as a reference  when available.

When a normal system  is available, 
we  host the corresponding software stack in the \texttt{X-Ray} VM and
 build the  correlation tree under the same bug-triggering workload.
For clarity, we name the tree from the normal system execution as the 
{\it reference tree}, 
which essentially  captures  the implicit rules 
among events in the normal  execution.
By comparing the trees, divergences that cause different symptoms
can be identified quickly.

\vspace{-0.1in}

\section{Preliminary Results}
\label{sec:experiments}
\vspace{-0.1in}


\begin{figure} 
    \begin{center}
           \includegraphics[width=3.1in]{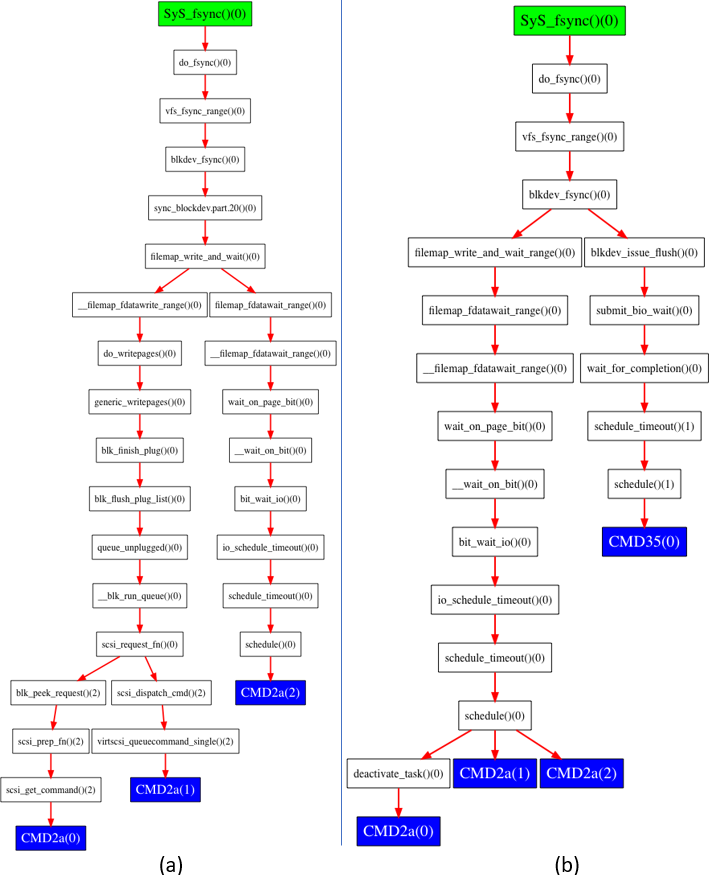}
    \end{center}
    \vspace{-0.3in}
    \caption{
        {\bf Comparison.}
      {\it  (a) the critical path from Figure~\ref{fig:crosslayergraph}; (b) the critical path from a reference tree.
    }
    }
    \vspace{-0.2in}
    \label{fig:comparison}
\end{figure}

We have built a preliminary prototype of 
{X-Ray} 
and applied it to 
diagnose 5 failure cases based on real bugs from the 
literature~\cite{Zheng-TOCS16-SSD,Lu-FAST13-FS,vjay-osdi18-blackboxtesting,SOSP19-filesystemsfuzzing}.
We discuss one case in details 
 and summarize the  results at the end.
 
 \smallskip
 \noindent
 {\bf Case Study.}
Figure~\ref{fig:crosslayergraph} shows 
a partial correlation tree for diagnosing a 
failure  where synchronous writes appear to be committed out 
of order on a raw block device. 
The  tree starts with a syscall node (green),
which triggers 704 kernel functions (white nodes)
and three device commands (blue nodes).
The red lines shows the critical path and nodes selected by one simple rule:
``ancestors of device commands'' (Rule\#3). 
The tree itself is part of the  original tree (not shown)  
selected via another simple rule:
``syscalls with WRITE commands'' (Rule\#1).

Figure~\ref{fig:comparison} (a) shows a zoomed in version of 
the critical path and nodes from Figure~\ref{fig:crosslayergraph}.
We can easily see that the \texttt{fsync} syscall only generates three WRITE (0x2a) commands 
without explicitly sending  SYNC\_CACHE (0x35)  command to the device,
which is incorrect based on  POSIX. 
Further investigation  confirms that the root cause lies in 
the \texttt{blkdev\_fsync} node on the critical path.

When a normal system is available, X-Ray may help more.
Figure~\ref{fig:comparison} (b) shows the critical path    on
 a reference tree. 
 Apparently, the SYNC\_CACHE (0x35) command appears, 
 and a new function \texttt{blkdev\_issue\_flush} is involved.
By comparison, 
it is clear that the difference stems from the \texttt{blkdev\_fsync} node.
\smallskip
\noindent
{\bf Summary.}
Table~\ref{tab:summary} summarizes the result.
Besides Rule\#1 and Rule\#3, 
we define another  Rule\#2: 
``functions  between the syscall and commands''.
Table~\ref{tab:summary} shows the node count of the original tree
and the node counts after applying each rule. 
The bold cell  means the  root cause 
can be covered by the  nodes selected via the corresponding rule. 
Overall, the simple rules can effectively
narrow down the search space for root cause (0.06\% - 4.97\% of the original trees).
We are studying other failure patterns and developing more intelligent rules.

\begin{table}[tb]
    \begin{center}
    \begin{tabular}{c|c|c|c|c}
    \hline
  ID  & Original & Rule\#1  &  Rule\#2 & Rule\#3 \\
    \hline
    \hline
 1 &  11,353 &  704   & 571  & 30   \\
  &  ({\bf 100\%}) &  ({\bf 6.20\%}) &   ({\bf 5.03\%}) & ({\bf 0.26\%})  \\
 \hline
 2 &  34,083 &  697   & 328 & 22 \\
   & ({\bf 100\%}) &({\bf 2.05\%})     &({\bf 0.96\%})   &({\bf 0.06\%})   \\
   \hline
  3 & 24,355 &  1,254 & 1,210  & 15  \\
  &({\bf 100\%})   &({\bf 5.15\%})     &({\bf 4.97\%})   &(0.06\%)   \\
  \hline
  4 & 273,653 &  10,230  & 9,953 &  40  \\
  &({\bf 100\%})    &({\bf 3.74\%})     &(3.64\%)   &(0.01\%)   \\
  \hline
 5 & 284,618 &   5,621 &  5,549 & 50  \\
 &({\bf 100\%})    &({\bf 1.97\%})     &({\bf 1.95\%})   &(0.04\%)   \\
    \hline
    \end{tabular}
    \end{center}
    \vspace{-0.25in}
    \caption{
    {\bf Result Summary.}
    }
    \vspace{-0.25in}
    \label{tab:summary}
    \end{table}

\vspace{-0.1in}
\section{Related Work}
\label{sec:related}
\vspace{-0.1in}


\noindent
{\bf Analyzing Storage Devices.}
Many researchers have studied the behaviors of storage devices 
in depth, including both HDDs~\cite{Remzi-TOS08-Corruption,Bairavasundaram-SIGMETRICS07-Latent,gibsonthesis,krioukov2008parity,bianca07}
 and 
 SSDs~\cite{nitin08,cai2014neighbor,
 Grupp12,Grupp09,kurata2006impact,li2015much,
 Schroeder-FAST16-Flash,powercut11,FAST20-SSDReliablityEnterprise}.
%
For example, 
Maneas \etal~\cite{FAST20-SSDReliablityEnterprise}
study the reliability of 1.4 million SSDs deployed in NetApp RAID systems.
Generally, these studies provide valuable insights for
reasoning complex  system failures involving device,
which is complementary to X-Ray.

\noindent
{\bf Diagnosing Distributed Systems.}
Great  efforts have been made on tracing and analyzing
distributed systems
~\cite{SOSP03-PerformanceDebugging,OSDI04Magpie,HotCloud09Mochi,Kandula-SIGCOMM09-DetailedDiagnosis,NSDI11-DiagnosingPerformance,PBahl-MultiLevelDependencies-sigcomm07}. 
For example, Aguilera \etal~\cite{SOSP03-PerformanceDebugging}
trace   network messages 
and infer causal relationships and latencies 
to diagnose performance issues. 
Similar to X-Ray, 
these methods need to align traces. 
However, 
their algorithms typically make use of  unique features of network events 
(e.g., RPC Send/Receive pairs, IDs in message headers),
which are not  available for X-Ray.
On the other hand, 
some statistic based methods~\cite{Kandula-SIGCOMM09-DetailedDiagnosis} are  
potentially applicable
 when enough  traces are collected.
\noindent
{\bf Software Engineering.}
Many software engineering techniques have been proposed for diagnosing user-level programs 
(e.g., program slicing~\cite{Agrawal-IEEESoftware91-backtracking,Weiser-CACM82-SlicesDebugging,Zhang-ICSE03-DynamicSlicing},
delta debugging~\cite{delta_debug,hierarchicaldeltadebugging}, checkpoint/re-execution~\cite{Flashback-USENIX04,Triage-SOSP03}).
In general, applying them directly to  the storage stack 
remains challenging 
due to the complexity.
On the other hand, some high-level ideas are likely applicable.
For example, Sambasiva \etal~\cite{NSDI11-DiagnosingPerformance}
apply  delta debugging to 
compare request flows  to diagnose performance
problems  in Ursa Minor~\cite{UrsaMinor-FAST05},
similar to the reference tree  part of X-Ray. 

\newpage
\vspace{-0.1in}
\section{Discussion Topics Section}
\label{sec:discussion}
\vspace{-0.1in}

\noindent
{\bf Emulating Storage Devices.}
As mentioned in \S\ref{sec:design},
sophisticated SATA/NVMe SSDs have been emulated in QEMU VM
~\cite{vissim,femu}. 
Among others, such efforts are important for realizing the
VM-based full-stack tracing and diagnosis. 
However, we do have observed some limitations of existing emulated devices, which may affect the failure reproducing (and thus diagnosis) in VM.
For example,  advanced  features like the TRIM operation are not fully supported on VSSIM or FEMU yet, 
but the Algolia failure case~\cite{algolia_blog} 
requires a TRIM-capable device to manifest. 
As a result, we are not able to reproduce the Algolia failure   in 
 VM. 
Therefore, emulating storage devices precisely
would be  helpful for the X-Ray approach and/or
  failure analysis in general, in addition to the other well-known benefits~\cite{vissim,femu}.
We would like to discuss how to improve the emulation accuracy
under practical constraints (e.g., confidentiality).
    \smallskip
    \noindent
    {\bf Deriving  Rules.}
  The automation of X-Ray depends on the rules. 
  The current prototype hard-coded a number of simple rules 
  based on our preliminary study  and 
    domain knowledge, 
  which is limited.
  We would like to explore
   other  implicit rules in the storage stack with other domain experts.
  Also, we plan to 
  collect correlation trees from normal system executions and 
  apply machine learning algorithms to derive the potential rules. 
   We would like to discuss   the feasibility. 
   

    \smallskip
    \noindent
    {\bf Other Usages of X-Ray}.
    We envision that some other analysis 
    could be enabled by \texttt{X-Ray}.
    For example, 
   with precise latency and casual relationships among events, 
    we may identify the  paths that are critical for I/O performance, similar to the request flow analysis 
    in  distributed systems~\cite{NSDI11-DiagnosingPerformance}. 
    Another possibility is to  
    measure the write amplification 
    at different layers across the stack.
    We would like to discuss the opportunities. 
    
    \smallskip
\noindent
{\bf Other Challenges of Failure Diagnosis.}
There are other challenges that are not covered by X-Ray.
For example, X-Ray assumes that there is a bug-triggering workload that can reliably 
lead to the failure. 
In practice, deriving  bug-triggering workloads from  user workloads 
(which may be huge or inconvenient to share) 
is often tricky~\cite{long-discussion,discussion2}. 
We would like to discuss such challenges. 

\smallskip
\noindent
{\bf Sharing Failure Logs.}
The cross-layer approach  would be most effective
for diagnosing obscure failures that involve 
both  the OS kernel and the device~\cite{algolia_blog,Zheng-TOCS16-SSD}. 
Based on our communications with storage  practitioners,
such failures are not uncommon.
However, the details of such failures 
are usually unavailable to the public,
which limits the use cases that could shape the design of reliability tools like X-Ray.
The availability of
 detailed failure logs at scale 
 is critical for moving 
 similar research efforts  forward.
We would like to discuss how to improve log sharing given constraints (e.g., privacy).

\vspace{-0.15in}
\section{Acknowledgements}
\label{sec:acknowledge}
\vspace{-0.1in}

We thank the anonymous reviewers for their insightful comments and suggestions.
This work was supported in part by the National Science Foundation (NSF) under grants CNS-1566554/1855565 and CCF-1717630/1853714. Any opinions, findings, and conclusions or recommendations expressed in this material are those of the author(s) and do not necessarily reflect the views of
NSF.

\bibliographystyle{plain}
\bibliography{reference}


\end{document}